\documentclass[copyright,creativecommons]{eptcs}
\providecommand{\event}{WWV 2015} 
\usepackage{breakurl}             
\usepackage[T1]{fontenc}
\usepackage[pdftex]{graphicx}
\usepackage{graphicx,wrapfig,subfig}
\usepackage{listings}

\title{Using a Machine Learning Approach to Implement and Evaluate Product Line Features}

\author{Davide Bacciu
\institute{Dipartimento di Informatica\\ Universit\`a di Pisa}
\email{bacciu@di.unipi.it}
\and
Stefania Gnesi
\institute{Istituto di Scienza e \\ Tecnologie dell'Informazione ``A.Faedo"\\
   	Consiglio Nazionale delle Ricerche, ISTI-CNR\\ Pisa, Italy}
\email{stefania.gnesi@isti.cnr.it}
\and
Laura Semini
\institute{Dipartimento di Informatica\\ Universit\`a di Pisa}
\email{semini@di.unipi.it}
}
\def\titlerunning{Using a Machine Learning Approach to Evaluate Procuct Line Features}
\def\authorrunning{D. Bacciu, S. Gnesi \& L. Semini}
\begin{document}
\maketitle

\begin{abstract}

Bike-sharing systems are a means of smart transportation in urban environments with the benefit of a positive impact on urban mobility.
In this paper we are interested in studying and modeling the behavior of features that permit the end user to access, with her/his web browser, the status of the Bike-Sharing system. In particular, we address features able to make a prediction on the system state.
We propose to use a machine learning approach to analyze usage patterns and learn computational models of such features from logs of system usage.

On the one hand, machine learning methodologies provide a powerful and general means to implement a wide choice of predictive features.  On the other hand, trained machine learning models are provided with a measure of predictive performance that can be used as a metric to assess the cost-performance trade-off of the feature. This provides a principled way to assess the runtime behavior of different components before putting them into operation.

\end{abstract}

\section{Introduction}

Product line engineering provides a way to manage variability during the entire design process~\cite{van2001notion} and is an important means for identifying variability needs early on. In this context, a feature represents a \textit{`logical unit of behavior that is specified by a set of functional and quality requirements'}~\cite{Soltani2012}. A \textit{Feature Model} is a compact representation of the commonalities and variabilities of the system, expressed as mandatory and optional features. Variability is achieved through the selection of the features that will be present in the final product.

In \textit{attributed feature models}, quantitative, non-functional characteristics of features are captured by attributes that are assigned to each feature. The use of attributed feature models is specifically useful for the decision-making process~\cite{Soltani2012}, as each stakeholder can make decisions taking into consideration both the features and the characteristics of the final product.
A number of techniques allow the configuration of feature models based on both functional and non-functional requirements~\cite{Soltani2012,benavides2013automated,Kacper2013}.

In this paper,  we want to explore the possibility of applying a machine learning (ML) approach to implement features and, at the same time, evaluate them to derive meaningful values to fill the attributed feature model. We concentrate on predictive features that are able to analyse the current state and some historical data, and provide some information to the user.
More in general, the purpose of the analysis is to evaluate the features and their possible combinations to help a stakeholder in deciding which product of a line to deploy, making the best possible compromise between cost and usefulness. The stakeholder may be a client that wants to buy a product, or, like in this case, a supplier that wants to know, before deployment, if a feature prediction accuracy will be worth the cost.
Finally, also when the cost is not an issue, it is interesting to have an assessment of the features before putting them in operation, to be sure they are accurate.

We focus, to explain our ideas, on Bike-sharing systems (BSS), which are a sustainable means of smart transportation with a positive impact on urban mobility. The quantitative analysis of bike-sharing systems, seen as collective adaptive systems (CAS) is a case study of the European project QUANTICOL (http://www.quanticol.eu).  CAS consist of a large number of spatially distributed entities, which may be competing for shared resources even when collaborating to reach common goals. The importance of the CAS in the context of urban mobility and in achieving societal goals means that it is necessary to carry out comprehensive analysis of their design and to investigate all aspects of their behavior before they are put into operation.

In previous work~\cite{TerBeek2014,TerBeekQC2014,TerBeek2013}, this case study was presented  and defined a discrete feature model, specifying several kinds of nonfunctional quantitative properties and behavioral characteristics. In particular, in~\cite{TerBeek2013} we established a chain of tools, each used to model a different aspect of the system, from feature modeling to product derivation and from quantitative evaluation of the attributes of products to model checking value-passing modal specifications.

This paper puts forward the idea of using Machine Learning (ML) methodologies to learn computational models of the features from BSS usage data. Throughout these methodologies, it is possible to learn the (unknown) relationship between the feature and its inputs by exploiting historical data representing examples of such input-output map. A trained model can then be used to provide predictions on future values of the feature in response to new input information, i.e. providing an implementation of the feature component. The advantage of such an approach is twofold: on the one hand, such methodologies provide a powerful and general means to realize a wide choice of predictive features for which there exist sufficient (and significative) historical data. On the other hand, trained ML models are provided with a measure of predictive performance that can be used as a metric to assess the cost-performance trade-off of the feature. This provides a principled way to assess the runtime behavior of different components before putting then into operation

\section{Case Study}

Many cities are currently adopting fully automated
public bike-sharing systems (BSS) as a green urban mode of transportation.
The concept is simple: a user arrives at a station, takes a bike, uses it for
a while and returns it to another station. A BSS can be conveniently considered and designed as a product line.

The automation of these systems permits to monitor the stations, to control if borrowed bikes are returned, to let the user pay for the usage, etc. In particular, a basic service of the system keeps track of all bikes and maintains a complete picture of which bikes are docked to each station and which ones are currently hired. For hired bikes, the system keeps track of the user name and departure station and time.

A user is interested in knowing the status of some station. In the case of the station being empty, the system may make a prediction  and infer if there is a bike that will be returned to that station soon.

\begin{figure}[!htp]
	\centering
        \includegraphics[width=0.5\textwidth]{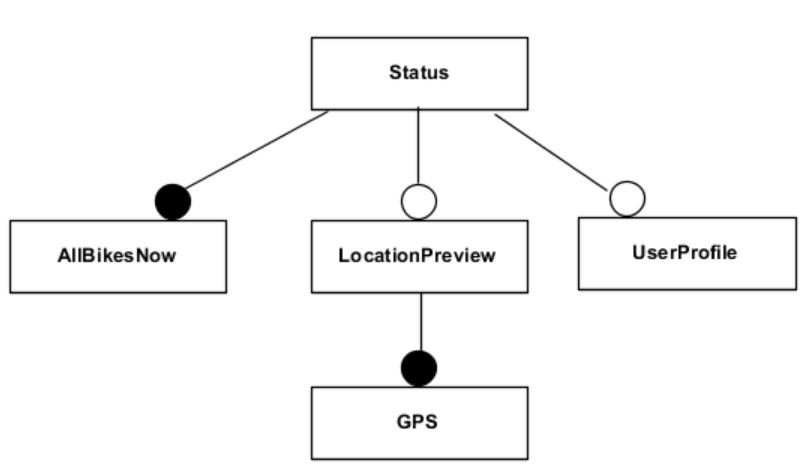}
	\caption{Status subsystem: the Feature Model}
	\label{fig:FeatMod}
\end{figure}

We can describe the feature model of the {\em Status} subsystem taking into account the above condition and hence it comprises a mandatory feature (for the basic service) and two optional ones, that can represent two different ways to predict the arrival of a bike, as shown in Figure~\ref{fig:FeatMod}.

Feature {\em AllBikesNow}, mandatory in each product of the line, keeps updated the current status of the service and tells how many bikes are parked in each station. This is used by the Bike-Sharing system administrators to know if some stations are empty or full and bikes need to be redistributed. The final user can access the status using their web browser before going to the station.

The {\em LocationPreview} feature predicts if a bike is going to arrive at a given station, and estimates the needed time. It makes use of a GPS to locate the bike: knowing the departure station and the path so far, the feature can predict the probability it will arrive at a station of interest in the next few minutes, and calculate the expected arrival time. Learning models are trained using historical traces of the BSS system usage.

Feature {\em UserProfile} also offers the same kind of prediction, but uses different data: the log, for each user, of all the uses of the Bike-Sharing system. For each use of the system, the log contains: departure time and station; arrival time and station. Analysing these data, {\em UserProfile} can predict if one of the bikes currently in use will arrive at the station of interest. Again, the feature returns a probability and the expected arrival time.

 In the next sections we provide a brief overview of ML and we discuss how this methodologies can be exploited to realize the {\em LocationPreview} and  {\em UserProfile} features of our BSS use case. Further, we summarize the main ideas underlying performance assessment in ML models.

\section{An Introduction to Machine Learning}

Machine learning provides computational models and methodologies to realize data-driven adaptive approaches to data analysis, pattern discovery and recognition, as well as to the predictive modeling of input-output data relationships. The term data-driven refers to the fact that ML approaches rely on (numerical) information encoded in the data, which is typically vectorial (i.e. multivariate data in a vector space) but can also be of relational type (i.e. compound information with a graph-based representation where edges encode relationships between the atomic information pieces) \cite{srl}.

ML is an active and wide research field comprising several paradigms, e.g. neural-inspired, probabilistic, kernel-based approaches, and addressing a variety of computational learning task types. For the purpose of product line feature modeling and evaluation, we focus on ML models and algorithms targeted at solving {\em supervised learning tasks}. Supervised learning refers to a specific class of ML problems that comprise learning of an (unknown) map $M: \mathcal{X} \rightarrow \mathcal{Y}$ between input information $x \in \mathcal{X}$ (e.g. a vector of attributes) and an output prediction $y \in \mathcal{Y}$ (in general, a vector of different dimensionality with respect to the input). Such an unknown map is learned from couples $\mathcal{D} = \{(x_1,y_1),\dots,(x_N,y_N)\}$ of input-output data, referred to as {\em training examples}, following a numerical routine targeted at the optimization of an error/performance function $E(\mathcal{D})$ which measures the quality of the predictions generated by the ML model.

ML models are characterized by two operational phases. The first is the {\em training} (or {\em learning}) phase, where ground-truth teaching information (encoded in training samples) is used to adapt the parameters regulating the response of the model so that its error (performance) $E(\mathcal{D})$ is reduced (increased, respectively). The {\em testing} (or {\em prediction}) phase, instead, supplies a trained model with novel input information (typically unseen at training time) to generate run-time predictions (i.e. to compute the learned map on novel data). The two phases are not always disjoint: {\em incremental learning} approaches exist that allow to continuously adapt the parameters of a ML model while this keeps providing its predictions in response to new input data.

In general, the final quality of the ML model predictions is influenced, on the one hand, by the quality of the training data, which should represent a sufficient and significative sample of the relationship to be modeled, and, on the other hand, by the adequacy of the learning model for the specific computational learning task. In this sense, different tasks, associated with different features to be modeled, may require to use learning models with different capabilities: in the following section, we analyze the nature of the tasks associated with BSS features prediction and we discuss which ML approaches are best suited to address them.

\section{Machine Learning for BSS Features}

Supervised learning approaches can be used to address modeling of product line features in our BSS scenario using logs of previous bike usage as training samples for the ML model. Here, we focus on the realization of the {\em UserProfile} and {\em LocationPreview} features. These two features are paradigmatic of two classes of learning tasks which require learning models of different nature and capabilities, i.e. static models for vectorial data and dynamic models for sequential data.

\subsection{User Profile}

The {\em UserProfile} feature requires to predict the destination station and arrival time of a bike given information on its pickup details and having knowledge of the BSS usage of the person who has picked-up the bike. A ML approach to realize such feature requires to train a different ML model for each user, using its personal usage logs as training data. In other words, a training dataset contains vector couples $(x_n,y_n)$ where the input attributes $x_n$ are a numerical encoding of the departure time and station, while the output $y_n$ encodes the associated time with destination and arrival station. At run-time, the feature prediction will be obtained by selecting the trained learning model associated with the user who has picked-up the bike and supplying it with the details of pickup time and station. Training of the learning models can also be performed at run-time: for instance, when a new customer subscribes to the service, the system starts collecting his/her usage information; as soon as sufficient usage data is collected, it is used to train a new learning model specific for the usage patterns of the customer. The same approach can be used to maintain the knowledge encoded in an existing customer model up-to-date: new examples are added to the log as the customer uses the system and a re-training of the learning model is performed as soon as a sufficient amount of new data is collected.

The prediction of the arrival station is an instance of a {\em classification problem} whose objective is to assign an input pattern to one of $K$ different and finite classes (bike stations, in our case study). The prediction of the time to destination, on the other hand, is an example of a {\em regression task}, where we are required to predict a generic (possibly continuous valued) output in response to the input pattern. These two problems can be more effectively solved by resorting to two separate and specialized learning models. The prediction of a probability estimate of class membership in place of a hard class assignment can be easily achieved by using a one-of-$K$ encoding of the classifier output coupled with a {\em soft-max} between the $K$ outputs. The one-of-$K$ encoding represents the fact that the $n$-th sample belongs to class $k$ (out of $K$) by a $K$ dimensional output vector $y_k$ having zeros on all components except for its $k$-th element $y_n(k)$ which is set to one. As a result, a trained learning model provided with an input $x$ at testing phase will produce a $K$-dimensional prediction $\tilde{y}$: the corresponding soft-max output $\bar{y}$ will again be a $K$-dimensional vector whose $k$-th element is
\[
    \bar{y}(k) = \frac{\tilde{y}(k)}{\sum_{l = 1}^K \tilde{y}(l)}.
\]

Data involved in the {\em UserProfile} feature is of static type, that is each training sample is a pair of identically and independently distributed vectors. The majority of the learning models in the literature have been designed to deal with such static vectorial data. For the purpose of implementing the {\em UserProfile} feature it is worth mentioning {\em Support Vector Machines} (SVMs), a family of supervised learning models which construct separating hyperplanes between the training vectors and exploit them to perform classification and regression \cite{cristSVM}. SVMs build on the concept of a linear separator (i.e. the separating hyperplane) and extend it to deal with non-linear problem by exploiting the so-called {\em kernel trick}, that is an implicit map of the input vector into an high-dimensional feature space by means of a non-linear map induced by a kernel function. SVM are highly effective classifiers and regressors for a wide-class of learning problems and several stable implementations are freely available \cite{svmlight,libsvm,svmTorch}. SVM training can be computationally demanding due to hyperparameters search, which can be a limiting factor for their use in run-time training. Further, it is difficult to interpret the result of a trained SVM. When interpretation of the results is an issue, probabilistic learning models found wide application: Naive Bayes and logistic regression are popular approaches \cite{nb}, although based on strong probabilistic assumptions which can be relaxed by resorting to more general Bayesian Networks \cite{Pearl:2000}.

\subsection{Location Preview}

The {\em LocationPreview} feature predicts the same output as the {\em UserProfile} feature using different input data, that are GPS trajectories corresponding to journeys performed by the BSS users. Trajectory data encodes a form of dynamical information of different nature with respect to the static vectorial data in {\em UserProfile}, requiring a radically different ML approach. A GPS trajectory is a form of sequential data, a type of structured information where the observation at a given point of the sequence is dependent on the context provided by the preceding or succeeding elements of the sequence. Such contextual information plays a role also in the learning task where, for instance, the decision on which will be the arrival station corresponding to a GPS trajectory cannot be taken based on the observation of a single element of the sequence, but should rather take into account the context provided by the full sequence or by a part of it. This requires learning models that can take into consideration such contextual information when computing their predictions, that are ML models for sequential/timeseries data. A straightforward approach to the problem is to use models for static data (such as those seen for the {\em UserProfile} feature) feeding them with a fixed-size chunk of the input sequence. This window of observations can be slid across the full length of the sequence, providing a prediction for each sequence element that can take into consideration the surrounding elements up to the window length. The key issue of such an approach is how to determine the correct size of the window for each learning problem. To address this issue, learning models have been proposed that are capable of maintaining a memory of the history of the input signals and to use it to compute their predictions. {\em Recurrent Neural Networks} (RNNs) \cite{rnn} are ML models that have been proposed specifically to deal with the dynamics of sequential information. They extended the original artificial neural networks paradigm with feedback connections that introduce a dynamic memory of the neuron activation which can be used to encode short to long term dependencies among the elements of the sequence, depending on the specific network architecture. In this context, the use of Reservoir Computing (RC) \cite{rc} has gained increasing interest as a modeling method for RNN, due to its ability  in conjugating  computational efficiency with the RNN capability of dealing with learning in temporal sequence domains. The underlying idea of the RC approach is to use a layer of sparsely connected recurrent neuron whose connections are initialized and left untrained; adaptation of the neural weights is restricted to the layer of output neurons. This allows to considerably reduce the computational complexity of training, which is a key issue if this is performed at run-time. RC models appear well suited for the implementation of the {\em LocationPreview} feature: in particular, they have already shown considerable efficacy in closely related learning tasks, such as the prediction of the destination room of trajectories of users walking in indoor environments \cite{esn}.

\subsection{Discussion}
The work reported in  \cite{miningLDA} applies ML to BSS data but it focuses mainly on mining usage models of BSS with the aim of identifying template behaviors which can be used as demand profiles for system management. In contrast, we propose to use ML as a modeling tool to build system features that can be used at run-time by the user application. At the same time, as explained in the next section, we propose to use the learning model error functions as part of the pre-deployment analysis to assess the cost-performance trade-off of the features to be included in the final BSS deployment. Finally, we also take into account the dynamic nature of trajectory data by using appropriate ML models, such as the RC approach, instead of adapting static models to perform spatio-temporal data analysis \cite{giot2014predicting}

\section{Performance evaluation of ML models}
A key aspect of ML models is the assessment of their predictive performance.
Good ML practice envisages a three-step process to build effective predictive learning models and reliably assess their performance.
\begin{enumerate}
  \item Training, which consists in adapting the parameters of the learning models using training data and numerical routines that optimize the model performance function (error).
  \item Model selection, which consists in estimating performance achieved by different learning models, including different hyper-parameter settings (i.e. model-tuning parameters set by the developer), in order to select the best model (with respect to the performance function)
  \item Final assessment, which consists in evaluating the performance of the selected model on new data, providing a measure of the generalization performance of the ultimately chosen model.
\end{enumerate}

Clearly, the latter step can be interpreted as a robust estimation of the performance of the feature implemented by the learning model when deployed in the run-time system. As such, it can be used as part of the product line to straightforwardly assess the efficiency-cost trade-off of the features implemented by ML models. Note that such an evaluation step can, in principle, exploit data logged by an existing system deployed by another client and different from the one being developed. Clearly, such estimated performance will provide an indication which will resemble the actual deployment only if the usage data available is coherent with the expected usage of the system under development. For  instance, it has to be expected that trajectory data for cities with considerably different BSS scales and topologies will not provide an adequate ground for comparison.

The three steps above can be implemented throughout a cross-validation scheme. The popular $K$-fold cross-validation would partition the available usage data into $K$ equally sized subsets, using $K-1$ groups for the first step while using the hold-out subset to assess the model performance (second step). This procedure is repeated for each of the $K$ possible choices of the held out group and the performance is then averaged over such $K$ choices. In the simplest scheme, this latter performance is used as final assessment of the model (third step). However, when key model selection choices are required in the second step, these are taken on the $K$-fold averaged performance, while the final assessment is computed on a completely external test set of hold-out data never used in the $K$-fold process. The actual form of the performance measure depends on the nature of the learning task, but it typically evaluates the discrepancy between the output predicted by the learning models and the desired (ground-truth) output. The Mean Absolute Error (MAE) is a popular choice to estimate the performance in regression task as the absolute value of the difference between the model output and the expected target output, averaged over the number of samples under consideration. For classification tasks, performance is often assessed  as class accuracy
\[
    acc_i = \frac{TP_i + TN_i}{N_i}
\]
where $N_i$ is the number of samples in the $i$-th class, while $TP_i$ and $TN_i$ are the number of true positive and true negative classifications predicted by the model for the $i$-th class.


\section{Conclusions}

We have discussed how a machine learning approach can be used to both implement and evaluate predictive Product Line Features. We addressed the case study of the European project QUANTICOL, concerning the quantitative analysis of bike-sharing systems (BSS). The features required by the case study are paradigmatic of two classes of learning tasks which require learning models of different nature and capabilities, i.e. static models for vectorial data and dynamic models for sequential data.

Such models are trained on historical usage data to realize a deployable implementation of the feature. In addition to that, a trained computational learning model is characterized by a measure of predictive performance that can be used to assess the cost-performance trade-off of the feature before putting it into operation.

We are currently training and validating a learning model for the {\sl UserProfile} feature using real-world usage data comprising more than 280.000 entries on the form

$$\langle UserID, \; leave \; station, \; leave \; date  \; and  \; time,  \;   return \; station, \; return \; date  \; and  \; time \rangle$$

covering all hires in Pisa across two years. As concerns the {\sl LocationPreview} feature, since bikes in Pisa are not equipped with GPS, we will look for data from different towns, which may in any case provide  a measure of predictive performance and let the stakeholders assess whether it is worth buying GPS trackers. A different solution is to use data coming from a simulation.

A general question is indeed associated with the concept of continuous learning, that is deciding when to activate model training and how to keep the feature up to date with respect to the availability of new usage data. Such choices can have an impact on the  prediction accuracy as well as on the stability of the learning model. For this specific aspect, we will initially rely on expert knowledge, but we will also explore possible automated decision processes.

To assess the cost-performance trade-off of the features, we plan to use Clafer, a general-purpose modeling language designed to represent domains, meta-models, components and variability models, like Feature models. Clafer has already been applied for modeling and optimization of product lines ~\cite{Kacper2013}~\cite{Antkiewicz2013}~\cite{Murashkin2013}.

\section*{Acknowledgments}
This research has been partly supported by the EU FP7-ICT FET-Proactive project QUANTICOL (600708) and by the Italian MIUR project CINA (PRIN 2010LHT4KM).

\bibliographystyle{eptcs}
\bibliography{bibliography}

\begin{thebibliography}{10}
\providecommand{\bibitemdeclare}[2]{}
\providecommand{\surnamestart}{}
\providecommand{\surnameend}{}
\providecommand{\urlprefix}{Available at }
\providecommand{\url}[1]{\texttt{#1}}
\providecommand{\href}[2]{\texttt{#2}}
\providecommand{\urlalt}[2]{\href{#1}{#2}}
\providecommand{\doi}[1]{doi:\urlalt{http://dx.doi.org/#1}{#1}}
\providecommand{\bibinfo}[2]{#2}

\bibitemdeclare{conference}{Antkiewicz2013}
\bibitem{Antkiewicz2013}
\bibinfo{author}{Micha\l \surnamestart Antkiewicz\surnameend},
  \bibinfo{author}{Kacper \surnamestart B\k{a}k\surnameend},
  \bibinfo{author}{Alexandr \surnamestart Murashkin\surnameend},
  \bibinfo{author}{Rafael \surnamestart Olaechea\surnameend},
  \bibinfo{author}{Jia Hui~(Jimmy) \surnamestart Liang\surnameend} \&
  \bibinfo{author}{Krzysztof \surnamestart Czarnecki\surnameend}
  (\bibinfo{year}{2013}): \emph{\bibinfo{title}{Clafer Tools for Product Line
  Engineering}}.
\newblock In: {\sl \bibinfo{booktitle}{Proceedings of the 17th International
  Software Product Line Conference Co-located Workshops}},
  \bibinfo{series}{SPLC '13 Workshops}, \bibinfo{publisher}{ACM}, pp.
  \bibinfo{pages}{130--135}, \doi{10.1145/2499777.2499779}.

\bibitemdeclare{article}{esn}
\bibitem{esn}
\bibinfo{author}{Davide \surnamestart Bacciu\surnameend},
  \bibinfo{author}{Paolo \surnamestart Barsocchi\surnameend},
  \bibinfo{author}{Stefano \surnamestart Chessa\surnameend},
  \bibinfo{author}{Claudio \surnamestart Gallicchio\surnameend} \&
  \bibinfo{author}{Alessio \surnamestart Micheli\surnameend}
  (\bibinfo{year}{2014}): \emph{\bibinfo{title}{An experimental
  characterization of reservoir computing in ambient assisted living
  applications}}.
\newblock {\sl \bibinfo{journal}{Neural Computing and Applications}}
  \bibinfo{volume}{24}(\bibinfo{number}{6}), pp. \bibinfo{pages}{1451--1464},
  \doi{10.1007/s00521-013-1364-4}.

\bibitemdeclare{inproceedings}{TerBeek2014}
\bibitem{TerBeek2014}
\bibinfo{author}{M.~H. \surnamestart ter Beek\surnameend},
  \bibinfo{author}{A.~\surnamestart Fantechi\surnameend} \&
  \bibinfo{author}{S.~\surnamestart Gnesi\surnameend} (\bibinfo{year}{2014}):
  \emph{\bibinfo{title}{{Challenges in Modelling and Analyzing Quantitative
  Aspects of Bike-Sharing Systems}}}.
\newblock In \bibinfo{editor}{T.~\surnamestart Margaria\surnameend} \&
  \bibinfo{editor}{B.~\surnamestart Steffen\surnameend}, editors: {\sl
  \bibinfo{booktitle}{Proceedings of the 6th International Symposium on
  Leveraging Applications of Formal Methods, Verification and Validation
  (ISoLA'14)}}, {\sl \bibinfo{series}{LNCS}} \bibinfo{volume}{8802},
  \bibinfo{publisher}{Springer}, pp. \bibinfo{pages}{351--367},
  \doi{10.1007/978-3-662-45234-9\_25}.

\bibitemdeclare{techreport}{TerBeekQC2014}
\bibitem{TerBeekQC2014}
\bibinfo{author}{M.~H. \surnamestart ter Beek\surnameend},
  \bibinfo{author}{A.~\surnamestart Fantechi\surnameend},
  \bibinfo{author}{S.~\surnamestart Gnesi\surnameend} \&
  \bibinfo{author}{F.~\surnamestart Mazzanti\surnameend}
  (\bibinfo{year}{2014}): \emph{\bibinfo{title}{{A collection of models of a
  bike-sharing case study}}}.
\newblock \bibinfo{type}{Technical Report} \bibinfo{number}{TR-QC-07-2014}.

\bibitemdeclare{techreport}{TerBeek2013}
\bibitem{TerBeek2013}
\bibinfo{author}{M.~H. \surnamestart ter Beek\surnameend},
  \bibinfo{author}{S.~\surnamestart Gnesi\surnameend} \&
  \bibinfo{author}{A.~\surnamestart Fantechi\surnameend}
  (\bibinfo{year}{2013}): \emph{\bibinfo{title}{{Chaining available tools to
  support the modelling and analysis of a bike-sharing product line: An
  experience report}}}.
\newblock \bibinfo{type}{Technical Report} \bibinfo{number}{TR-QC-02-2013}.

\bibitemdeclare{incollection}{benavides2013automated}
\bibitem{benavides2013automated}
\bibinfo{author}{David \surnamestart Benavides\surnameend},
  \bibinfo{author}{Pablo \surnamestart Trinidad\surnameend} \&
  \bibinfo{author}{Antonio \surnamestart Ruiz-Cort{\'e}s\surnameend}
  (\bibinfo{year}{2013}): \emph{\bibinfo{title}{Automated Reasoning on Feature
  Models}}.
\newblock In: {\sl \bibinfo{booktitle}{Seminal Contributions to Information
  Systems Engineering}}, \bibinfo{publisher}{Springer}, pp.
  \bibinfo{pages}{361--373}, \doi{10.1007/11431855\_34}.

\bibitemdeclare{phdthesis}{Kacper2013}
\bibitem{Kacper2013}
\bibinfo{author}{Kacper \surnamestart B\k{a}k\surnameend}
  (\bibinfo{year}{2013}): \emph{\bibinfo{title}{Modeling and Analysis of
  Software Product Line Variability in Clafer}}.
\newblock Ph.D. thesis, \bibinfo{school}{University of Waterloo}.

\bibitemdeclare{article}{libsvm}
\bibitem{libsvm}
\bibinfo{author}{Chih-Chung \surnamestart Chang\surnameend} \&
  \bibinfo{author}{Chih-Jen \surnamestart Lin\surnameend}
  (\bibinfo{year}{2011}): \emph{\bibinfo{title}{{LIBSVM}: A library for support
  vector machines}}.
\newblock {\sl \bibinfo{journal}{ACM Transactions on Intelligent Systems and
  Technology}} \bibinfo{volume}{2}, pp. \bibinfo{pages}{27:1--27:27},
  \doi{10.1145/1961189.1961199}.
\newblock \bibinfo{note}{Software available at
  \url{http://www.csie.ntu.edu.tw/~cjlin/libsvm}}.

\bibitemdeclare{article}{svmTorch}
\bibitem{svmTorch}
\bibinfo{author}{R.~\surnamestart Collobert\surnameend} \&
  \bibinfo{author}{S.~\surnamestart Bengio\surnameend} (\bibinfo{year}{2001}):
  \emph{\bibinfo{title}{{SVMT}orch: Support Vector Machines for Large-Scale
  Regression Problems}}.
\newblock {\sl \bibinfo{journal}{Journal of Machine Learning Research}}
  \bibinfo{volume}{1}, pp. \bibinfo{pages}{143--160},
  \doi{10.1162/15324430152733142}.

\bibitemdeclare{inproceedings}{miningLDA}
\bibitem{miningLDA}
\bibinfo{author}{Etienne \surnamestart Come\surnameend},
  \bibinfo{author}{Njato~Andry \surnamestart Randriamanamihaga\surnameend},
  \bibinfo{author}{Latifa \surnamestart Oukhellou\surnameend} \&
  \bibinfo{author}{Patrice \surnamestart Aknin\surnameend}
  (\bibinfo{year}{2014}): \emph{\bibinfo{title}{{Spatio-temporal Analysis of
  Dynamic Origin-Destination Data Using Latent Dirichlet Allocation:
  Application to V{\'e}lib' Bike Sharing System of Paris}}}.
\newblock In: {\sl \bibinfo{booktitle}{{TRB 93rd Annual meeting}}},
  \bibinfo{address}{France}.
\newblock \urlprefix\url{https://hal.archives-ouvertes.fr/hal-01052951}.

\bibitemdeclare{book}{srl}
\bibitem{srl}
\bibinfo{author}{Lise \surnamestart Getoor\surnameend} \& \bibinfo{author}{Ben
  \surnamestart Taskar\surnameend} (\bibinfo{year}{2007}):
  \emph{\bibinfo{title}{Introduction to Statistical Relational Learning
  (Adaptive Computation and Machine Learning)}}.
\newblock \bibinfo{publisher}{The MIT Press}.

\bibitemdeclare{inproceedings}{giot2014predicting}
\bibitem{giot2014predicting}
\bibinfo{author}{Romain \surnamestart Giot\surnameend} \&
  \bibinfo{author}{Rapha{\"e}l \surnamestart Cherrier\surnameend}
  (\bibinfo{year}{2014}): \emph{\bibinfo{title}{Predicting bikeshare system
  usage up to one day ahead}}.
\newblock In: {\sl \bibinfo{booktitle}{Computational Intelligence in Vehicles
  and Transportation Systems (CIVTS), 2014 IEEE Symposium on}},
  \bibinfo{organization}{IEEE}, pp. \bibinfo{pages}{22--29},
  \doi{10.1109/CIVTS.2014.7009473}.

\bibitemdeclare{incollection}{svmlight}
\bibitem{svmlight}
\bibinfo{author}{T.~\surnamestart Joachims\surnameend} (\bibinfo{year}{1999}):
  \emph{\bibinfo{title}{Making large-Scale {SVM} Learning Practical}}.
\newblock In \bibinfo{editor}{B.~\surnamestart Sch{\"o}lkopf\surnameend},
  \bibinfo{editor}{C.~\surnamestart Burges\surnameend} \&
  \bibinfo{editor}{A.~\surnamestart Smola\surnameend}, editors: {\sl
  \bibinfo{booktitle}{Advances in Kernel Methods - Support Vector Learning}},
  chapter~\bibinfo{chapter}{11}, \bibinfo{publisher}{MIT Press},
  \bibinfo{address}{Cambridge, MA}, pp. \bibinfo{pages}{169--184}.

\bibitemdeclare{book}{rnn}
\bibitem{rnn}
\bibinfo{author}{Stefan~C. \surnamestart Kremer\surnameend}
  (\bibinfo{year}{2001}): \emph{\bibinfo{title}{Field Guide to Dynamical
  Recurrent Networks}}.
\newblock \bibinfo{publisher}{Wiley-IEEE Press},
  \doi{10.1109/9780470544037.fmatter}.

\bibitemdeclare{article}{rc}
\bibitem{rc}
\bibinfo{author}{Mantas \surnamestart Luko\v{s}evi\v{c}ius\surnameend} \&
  \bibinfo{author}{Herbert \surnamestart Jaeger\surnameend}
  (\bibinfo{year}{2009}): \emph{\bibinfo{title}{Reservoir Computing Approaches
  to Recurrent Neural Network Training}}.
\newblock {\sl \bibinfo{journal}{Comput. Sci. Rev.}}
  \bibinfo{volume}{3}(\bibinfo{number}{3}), pp. \bibinfo{pages}{127--149},
  \doi{10.1016/j.cosrev.2009.03.005}.

\bibitemdeclare{conference}{Murashkin2013}
\bibitem{Murashkin2013}
\bibinfo{author}{Alexandr \surnamestart Murashkin\surnameend},
  \bibinfo{author}{Micha{\l} \surnamestart Antkiewicz\surnameend},
  \bibinfo{author}{Derek \surnamestart Rayside\surnameend} \&
  \bibinfo{author}{Krzysztof \surnamestart Czarnecki\surnameend}
  (\bibinfo{year}{2013}): \emph{\bibinfo{title}{Visualization and Exploration
  of Optimal Variants in Product Line Engineering}}.
\newblock In: {\sl \bibinfo{booktitle}{Software Product Line Conference}},
  \bibinfo{address}{Tokyo, Japan}, \doi{10.1145/2491627.2491647}.

\bibitemdeclare{incollection}{nb}
\bibitem{nb}
\bibinfo{author}{Andrew~Y. \surnamestart Ng\surnameend} \&
  \bibinfo{author}{Michael~I. \surnamestart Jordan\surnameend}
  (\bibinfo{year}{2002}): \emph{\bibinfo{title}{On Discriminative vs.
  Generative Classifiers: A comparison of logistic regression and naive
  Bayes}}.
\newblock In \bibinfo{editor}{T.G. \surnamestart Dietterich\surnameend},
  \bibinfo{editor}{S.~\surnamestart Becker\surnameend} \&
  \bibinfo{editor}{Z.~\surnamestart Ghahramani\surnameend}, editors: {\sl
  \bibinfo{booktitle}{Advances in Neural Information Processing Systems 14}},
  \bibinfo{publisher}{MIT Press}, pp. \bibinfo{pages}{841--848}.

\bibitemdeclare{book}{Pearl:2000}
\bibitem{Pearl:2000}
\bibinfo{author}{Judea \surnamestart Pearl\surnameend} (\bibinfo{year}{2000}):
  \emph{\bibinfo{title}{Causality: Models, Reasoning, and Inference}}.
\newblock \bibinfo{publisher}{Cambridge University Press},
  \bibinfo{address}{New York, NY, USA}.

\bibitemdeclare{book}{cristSVM}
\bibitem{cristSVM}
\bibinfo{author}{John \surnamestart Shawe-Taylor\surnameend} \&
  \bibinfo{author}{Nello \surnamestart Cristianini\surnameend}
  (\bibinfo{year}{2004}): \emph{\bibinfo{title}{Kernel Methods for Pattern
  Analysis}}.
\newblock \bibinfo{publisher}{Cambridge University Press},
  \bibinfo{address}{New York, NY, USA}, \doi{10.1017/CBO9780511809682}.

\bibitemdeclare{inproceedings}{Soltani2012}
\bibitem{Soltani2012}
\bibinfo{author}{Samaneh \surnamestart Soltani\surnameend},
  \bibinfo{author}{Mohsen \surnamestart Asadi\surnameend},
  \bibinfo{author}{Dragan \surnamestart Ga\v{s}evi\'{c}\surnameend},
  \bibinfo{author}{Marek \surnamestart Hatala\surnameend} \&
  \bibinfo{author}{Ebrahim \surnamestart Bagheri\surnameend}
  (\bibinfo{year}{2012}): \emph{\bibinfo{title}{Automated Planning for Feature
  Model Configuration Based on Functional and Non-functional Requirements}}.
\newblock In: {\sl \bibinfo{booktitle}{Proceedings of the 16th International
  Software Product Line Conference - Volume 1}}, \bibinfo{series}{SPLC '12},
  \bibinfo{publisher}{ACM}, \bibinfo{address}{New York, NY, USA}, pp.
  \bibinfo{pages}{56--65}, \doi{10.1145/2362536.2362548}.

\bibitemdeclare{inproceedings}{van2001notion}
\bibitem{van2001notion}
\bibinfo{author}{Jilles \surnamestart Van~Gurp\surnameend},
  \bibinfo{author}{Jan \surnamestart Bosch\surnameend} \&
  \bibinfo{author}{Mikael \surnamestart Svahnberg\surnameend}
  (\bibinfo{year}{2001}): \emph{\bibinfo{title}{On the notion of variability in
  software product lines}}.
\newblock In: {\sl \bibinfo{booktitle}{Software Architecture, 2001.
  Proceedings. Working IEEE/IFIP Conference on}}, \bibinfo{organization}{IEEE},
  pp. \bibinfo{pages}{45--54}, \doi{10.1109/WICSA.2001.948406}.

\end{thebibliography}

\end{document}